\documentclass{aa}

\usepackage{graphicx}
\usepackage{txfonts}
\usepackage{url}

\begin{document}

 \title{Origins of the water-rich atmosphere of the sub-Neptune GJ~9827~d}
\subtitle{Coupled magma ocean--atmosphere evolution favours a near-IW, carbon-poor mantle}
\titlerunning{Magma ocean origin of the GJ~9827~d atmosphere}

\author{Imre Kisv\'ardai\inst{1}
        \and
        Tim Lichtenberg\inst{1}
        \and
        Harrison Nicholls\inst{2}
        \and
        Caroline Piaulet-Ghorayeb\inst{3}}
\authorrunning{Kisv\'ardai et al.}

\institute{Kapteyn Astronomical Institute, University of Groningen,
           Landleven 12, 9747 AD Groningen, The Netherlands\\
           \email{i.g.kisvardai@rug.nl}
           \and
           Institute of Astronomy, University of Cambridge,
           Madingley Road, Cambridge CB3 0HA, United Kingdom
           \and
           Department of Astronomy \& Astrophysics, University of Chicago,
           5640 South Ellis Avenue, Chicago, IL 60637, USA}

\date{Received ; accepted }

\abstract
  {The sub-Neptune GJ~9827~d has a bulk density between that of rock and of a hydrogen-dominated envelope, and JWST/NIRISS transmission spectroscopy indicates a water-rich, high-mean-molecular-weight atmosphere with no detected carbon-bearing species. Whether such an atmosphere can arise from, and survive after, a magma-ocean stage is unclear.}
  {We test whether coupled interior--atmosphere evolution from a fully molten start reproduces the bulk density and atmospheric H$_2$O and CH$_4$ abundances of GJ~9827~d at its present age, and which initial conditions this requires.}
  {With the \textsc{PROTEUS} framework, we simulated planets spanning primordial hydrogen and carbon inventory, core mass fraction and mantle oxygen fugacity, coupling interior evolution, melt--gas partitioning, radiative--convective atmosphere, stellar evolution and energy-limited escape until 5.465~Gyr or atmospheric loss. Final states are compared with the bulk density and the minimum H$_2$O and maximum CH$_4$ volume mixing ratios.}
  {Nearly half of the completed runs retain an atmosphere to the present age. Mantle redox controls the outcome: at $f$O$_2 \le \Delta$IW$-2$ most runs lose their atmosphere and survivors keep H$_2$--CH$_4$ envelopes, whereas oxidised mantles ($\ge \Delta$IW$+2$) retain H$_2$O-dominated envelopes too dense to match the observations. Sixty-eight runs satisfy the density and water constraints, all at $\Delta$IW$\pm0$ or $+1$ with $\ge$6325~ppmw H. They end with bulk densities of 2.50--3.16~g\,cm$^{-3}$ and mixed atmospheres of on average 52\% H$_2$O, 45\% H$_2$ and 2\% CH$_4$. The CH$_4$ upper limit requires C/H~$\le 10^{-2}$. Several matching runs undergo \emph{reflation}, a $>$5\% density decrease driven by late outgassing of low-mean-molecular-weight species \citep{cesario2026}, which predicts 12--19\% atmospheric CH$_4$.}
  {A magma-ocean origin can explain GJ~9827~d if its mantle equilibrated near the iron--w\"ustite buffer and it formed with $\sim$1~wt\% H and C/H~$\le10^{-2}$ by mass. The atmosphere is a mixed H$_2$O--H$_2$ envelope rather than pure steam, consistent with JWST retrievals assuming an H$_2$/He background.}


\maketitle

\section{Introduction}
\label{sec:intro}

Planets with radii of about 1.5--4~$R_\oplus$ are common and within reach of current spectral characterisation, but have no counterparts in the Solar System \citep{lichtenberg2023, nicholls2026}. This regime contains the super-Earth and sub-Neptune populations, separated by a relative paucity of planets known as the radius valley. How the two populations originate remains unclear \citep{zhu2021, kempton2024}. Two scenarios are commonly considered \citep{lichtenberg2025, lichtenberg2023}: (1) gas dwarfs, in which both populations form with H$_2$-dominated envelopes and diverge through long-term atmospheric escape \citep{owen2017evaporation, gupta2019sculpting}; and (2) water worlds, in which super-Earths and sub-Neptunes differ in the amount of H$_2$O they acquired, depending on where they formed relative to the ice line \citep{aguichine2021mass, innes2023runaway, benneke2024jwst}. A mixture of both may explain the populations as a whole but does not settle the nature of any individual planet \citep{ rice2019masses, kosiarek2020physical, Venturini2024AA, burn2024radius, nicholls2026}.

GJ~9827 hosts three transiting planets \citep{niraula2017three, rodriguez2018system, prieto2018mass}. The inner planets b and c have bulk densities consistent with rocky compositions, whereas the outermost planet, d ($R = 1.89$--$1.98\,R_\oplus$; \citealt{passegger2024compact, piaulet2024jwst}), is markedly less dense, which implies a substantial volatile component \citep{teske2018magellan, prieto2018mass, passegger2024compact}. Interpretation of JWST/NIRISS transmission spectroscopy data has since revealed a water-rich atmosphere with a high mean molecular weight, together with upper limits on CH$_4$ and NH$_3$ \citep{piaulet2024jwst}. \citet{piaulet2024jwst} suggest that if all of its volatiles were held in a pure steam envelope, the planet would be $32\pm10$\% water by mass. Such water could derive from ice-rich building blocks or be produced by reactions between a hydrogen-rich envelope and a magma ocean \citep{lichtenberg2023, miozzi2025}, although the oxygen budget of the magma limits how much water the latter route can make \citep{kimura2026}. Escape observations point the same way: Ly$\alpha$, H$\alpha$ and He\,\textsc{i} transits show no extended exosphere \citep[see also][]{piaulet2024jwst}, which argues against a hydrogen-dominated envelope losing mass at the high rates expected for this irradiation \citep{kasper2020nondetection, carleo2021multiwavelength, krishnamurthy2023absence}. GJ~9827~d is therefore a benchmark for volatile-rich, highly irradiated small planets near the radius valley. What remains unclear is whether such an atmosphere can be a long-lived end state of a planet that formed hot and molten, or whether it requires a specific formation history.

A related question is whether the present-day density of such a planet reflects its long-term state at all. Coupled interior--atmosphere models predict that the radius of an irradiated planet need not shrink monotonically: when escape removes a heavy, carbon-rich secondary atmosphere from a reduced magma ocean, lighter hydrogen-bearing gas outgassed from the melt can replace it and transiently inflate the planet, a process termed reflation \citep{cesario2026}. A planet observed during such an episode would appear less dense than its long-term evolution implies, and coupled models suggest that even sub-Neptunes older than a gigayear may still be in a transient state \citep{nicholls2026}.

Here we connect plausible initial conditions of GJ~9827~d to its present-day bulk density and atmospheric water and methane abundances with a grid of coupled evolution models. We aim to determine the primordial composition of the interior and atmosphere and the present-day geophysical parameters of the interior that are consistent with the observed bulk density and spectral characteristics of GJ~9827~d. We use the \textsc{PROTEUS} framework\footnote{\url{https://proteus-framework.org}} \citep{lichtenberg2021, nicholls2024a, nicholls2024b}, which evolves the interior and atmosphere of a planet together from a fully molten start: interior structure and mantle crystallisation, melt--gas partitioning and speciation of C--H--N--O--S volatiles, radiative--convective energy transport through the atmosphere, stellar evolution, and escape. The grid spans the hydrogen inventory, the C/H ratio over six orders of magnitude, the core mass fraction and the mantle oxygen fugacity. Planets are thought to vary widely in their initial volatile budgets \citep{drazkowska2023, kriijt2023} and in their mantle redox state \citep{sossi2020, gaillard2021}, and redox controls both volatile solubility in silicate melt and the speciation of the outgassed atmosphere \citep{suer2023, sossi2023, nicholls2024a}. Section~\ref{sec:methods} describes the model and the grid, Sect.~\ref{sec:results} the results, Sect.~\ref{sec:discussion} their interpretation and limitations, and Sect.~\ref{sec:conclusions} our conclusions.

\section{Methods}
\label{sec:methods}

\textsc{PROTEUS}\footnote{\url{https://github.com/FormingWorlds/PROTEUS}} \citep{lichtenberg2021, nicholls2024a, nicholls2024b} couples separate modules for interior structure, interior energetics, outgassing, climate, stellar evolution and escape, which exchange boundary conditions at every coupling step. The framework is validated through continuous physical and numerical tests on the GitHub platform\footnote{\url{https://proteus-framework.org/validation}} and through the CHILI code intercomparison \citep{2026PSJ.....7..108L,2026arXiv260624757N}. Below we describe the configuration used here. Table~\ref{tab:fixed} lists the main fixed settings and Table~\ref{tab:params} the grid axes.

\subsection{Interior structure and thermal evolution}
\label{subsec:interior}

The interior structure is computed with \textsc{Zalmoxis}\footnote{\url{https://github.com/FormingWorlds/Zalmoxis}} \citep{lichtenberg2026shrink}, which solves for hydrostatic equilibrium of an iron core and a silicate mantle for a given total mass and core mass fraction, using the PALEOS equation-of-state tables for iron and MgSiO$_3$ \citep{Attia2026A&A}. In the grid presented here the structure is solved once, for the initial, fully molten state, and is kept fixed thereafter; the consequences of this choice are discussed in Sect.~\ref{subsec:disc-limits}.

The thermal evolution of the mantle is computed with \textsc{Aragog}\footnote{\url{https://github.com/FormingWorlds/aragog}} \citep{lichtenberg2026shrink}, a one-dimensional interior energetics model in the mixing length energetics tradition of \textsc{SPIDER} \citep{bower2018, bower2019, bower2022} that tracks the radial profiles of temperature, melt fraction and the associated material properties as the mantle crystallises from the bottom up. The mantle starts fully molten, with a temperature profile 100~K above the liquidus, consistent with canonical magma ocean initial conditions \citep{abe1986, hamano2015}. The transition from liquid-like to solid-like rheology is centred at a melt fraction of 0.4 with a width of 0.2 \citep[cf.][]{costa2009}; melt and solid viscosities are $10^{2}$ and $10^{22}$~Pa\,s. Radiogenic heating from $^{40}$K, U and Th is included at bulk-silicate-Earth abundances. Tidal heating, which can prolong magma oceans on close-in planets \citep{nicholls2025b}, is not included here.

\subsection{Volatile partitioning and outgassing}
\label{subsec:volatiles}

Volatile exchange between the magma ocean and the atmosphere is computed with the \textsc{CALLIOPE}\footnote{\url{https://github.com/FormingWorlds/CALLIOPE}} module \citep{bower2019, bower2022, nicholls2024a, nicholls2026}, which solves for gas-phase chemical equilibrium \citep{chase1998} together with mass balance and solubility in the melt. It considers ten species: H$_2$O, H$_2$, CO$_2$, CO, CH$_4$, N$_2$, NH$_3$, S$_2$, SO$_2$ and H$_2$S. The solubility laws are those of \citet{sossi2023} for H$_2$O, \citet{dixon1995} for CO$_2$, \citet{armstrong2015} for CO, \citet{ardia2013} for CH$_4$, \citet{dasgupta2022} for N$_2$ and \citet{gaillard2022} for sulfur. Other species are treated as insoluble. The IW buffer follows \citet{oneill2002}. Atmospheric partial pressures therefore depend on the mass of melt in the mantle, the surface gravity and temperature, and the oxygen fugacity at the surface \citep{bower2019, sossi2020, gaillard2021, maurice2024}. The oxygen fugacity is held at a constant offset from the iron--w\"ustite (IW) buffer, $\Delta$IW, throughout each run. Volatiles are not retained in the crystallising solid: when the mantle fully solidifies, the remaining dissolved inventory is released to the atmosphere. Graphite and other condensed carbon phases are not included.

The volatile inventory is specified by element. Hydrogen is given in ppm by weight (ppmw) relative to the mantle mass, carbon as a C/H mass ratio, and nitrogen and sulfur as fixed N/H and S/H mass ratios of 0.01. Oxygen is set by the chemistry of the initial equilibrium state.

\subsection{Atmosphere}
\label{subsec:atm}

The atmosphere is modelled with \textsc{AGNI} \citep{nicholls2025a, nicholls2024b}, which solves for the radiative--convective temperature--pressure structure and energy fluxes with a Newton--Raphson method, so that energy is conserved throughout the column within a small tolerance. Radiative transfer uses \textsc{SOCRATES} \citep{edwards1996, sergeev2023} with 48 correlated-$k$ bands \citep{lacis1991}. Gas opacities come from the DACE database \citep{grimm2021}, complemented by continuum absorption \citep{mlawer2023, nicholls2024a}, and Rayleigh scattering is included. Convection is treated with mixing-length theory under the Schwarzschild criterion \citep{vitense1953, robinson2014, joyce2023}, and condensation and rainout are included, but cloud and haze formation are not. The atmosphere follows the ideal-gas law with a well-mixed composition set by the outgassing calculation. The surface is a grey body with an albedo of 0.1. The column is resolved on 50 logarithmically spaced levels from the surface to $10^{-6}$~bar.

The model transit radius $R_\mathrm{obs}$ is taken at the level where the optical depth at 1.125~$\mu$m reaches 0.02, and the bulk density compared with observations is $\rho_\mathrm{obs} = 3M_\mathrm{p}/(4\pi R_\mathrm{obs}^3)$, where $M_\mathrm{p}$ is the instantaneous planet mass.

\subsection{Star, orbit and escape}
\label{subsec:star}

The host star, with an adopted mass of 0.606~$M_\odot$ (within the range of estimates given by \citealt{passegger2024compact}), is evolved with \textsc{MORS} \citep{johnstone2021active} on the stellar evolution tracks of \citet{spada2013}, starting at a stellar age of 0.1~Gyr. GJ~9827 has no observed panchromatic spectrum in the libraries available to us, so we use the MUSCLES spectrum of the K dwarf HD~85512 \citep{france2016, youngblood2016, loyd2016} as a stand-in. The planet orbits at 0.0563~AU \citep{passegger2024compact}. A zenith angle of 48.19$^\circ$ and an instellation scale factor of 0.375 give the global-mean instellation. Tidal evolution of the orbit is not modelled.

Atmospheric escape is computed with the \textsc{ZEPHYRUS}\footnote{\url{https://github.com/FormingWorlds/ZEPHYRUS}} module \citep{postolec2026}, using an energy-limited prescription \citep{watson1981, erkaev2007} with an efficiency of 0.1. The XUV flux follows the \textsc{MORS} activity evolution and is absorbed at the $5\times10^{-5}$~bar level, which is capped at the Hill radius. The escaping gas has the bulk composition of the outgassed reservoir, so escape does not fractionate the escaping elemental inventory.

\begin{table}
\caption{Fixed model settings.}
\label{tab:fixed}
\centering
\begin{tabular}{l l}
\hline\hline
Quantity & Value \\
\hline
Initial planet mass & 3.02~$M_\oplus$ \\
Orbital distance & 0.0563~AU \\
Stellar mass & 0.606~$M_\odot$ \\
Stellar spectrum & HD~85512 (MUSCLES) \\
Model start / end age & 0.1 / 5.465~Gyr \\
Mantle / core EOS & PALEOS MgSiO$_3$ / Fe \\
Initial mantle state & 100~K above liquidus \\
Radiogenic heating & K, U, Th at BSE values \\
N/H, S/H (mass) & 0.01, 0.01 \\
Correlated-$k$ bands & 48 \\
Surface albedo & 0.1 (grey) \\
Escape efficiency & 0.1 \\
Escape stop & $P_\mathrm{surf} < 3$~bar \\
\hline
\end{tabular}
\end{table}

\begin{table}
\caption{Simulation grid axes. All combinations were run (945 cases).}
\label{tab:params}
\centering
\begin{tabular}{l l}
\hline\hline
Parameter & Values \\
\hline
H (ppmw of mantle) & 1000, 2515, 6325, 15\,905, 40\,000 \\
Mantle $\Delta$IW & $-4, -3, \dots, +4$ \\
C/H (mass) & $10^{-6}, 10^{-5}, \dots, 10^{-1}, 1$ \\
Core mass fraction & 0.215, 0.27, 0.325 \\
\hline
\end{tabular}
\tablefoot{The H values are spaced evenly in $\log_{10}$ between $10^3$ and $4\times10^4$~ppmw. The C/H values are spaced by one decade.}
\end{table}

\begin{table*}
\caption{Observed properties of GJ~9827~d and their use in this work.}
\label{tab:obs}
\centering
\small
\setlength{\tabcolsep}{4pt}
\begin{tabular}{l l l l}
\hline\hline
Quantity & Value & Reference & Use \\
\hline
Mass, $M_\mathrm{p}$ & $3.02^{+0.58}_{-0.57}\,M_\oplus$ & \citet{passegger2024compact} & Initial planet mass \\
Radius, $R_\mathrm{p}$ & $1.89^{+0.16}_{-0.14}\,R_\oplus$ & \citet{passegger2024compact} & Comparison (Sect.~\ref{subsec:disc-mass}) \\
Bulk density, $\rho$ & $2.454\pm0.739$~g\,cm$^{-3}$ & \citet{passegger2024compact} & Constraint ($1\sigma$ interval) \\
Orbital distance & 0.0563~AU & \citet{passegger2024compact} & Fixed input \\
Stellar mass & 0.606~$M_\odot$ & \citet{passegger2024compact} & Fixed input \\
System age & $5.465\pm4.058$~Gyr & \citet{passegger2024compact} & End time (central value) \\
$\log_{10} X_\mathrm{H_2O}$ & $>-0.50$ / $>-1.86$ & \citet{piaulet2024jwst} & Constraint (agnostic value) \\
$\log_{10} X_\mathrm{CH_4}$ & $<-3.60$ / $<-3.71$ & \citet{piaulet2024jwst} & Third, weaker constraint \\
Mean molecular weight & $18.02^{+0.20}_{-0.61}$ / $9.84^{+4.51}_{-4.64}$~g\,mol$^{-1}$ & \citet{piaulet2024jwst} & Comparison \\
$\log_{10}$ C/H (mass) & $-2.8\pm2.0$ / $-2.9\pm2.0$ & \citet{piaulet2024jwst} & Comparison \\
\hline
\end{tabular}
\tablefoot{Where two values are given, they are from the retrievals of \citet{piaulet2024jwst} with an agnostic and an H$_2$/He background gas, respectively. Limits on mixing ratios are one-sided 95\% limits; two-sided intervals are 68\%.}
\end{table*}

\subsection{Grid, termination and observational constraints}
\label{subsec:sims}

The simulation grid axes (Table~\ref{tab:params}) cover the uncertain bulk composition and primordial volatile endowment of the planet. The ranges are motivated by planet formation models \citep{drazkowska2023, kriijt2023}, by geochemical constraints on Solar System bodies \citep{wadhwa2001, cartier2019}, and by the compositions of material accreted by polluted white dwarfs \citep{doyle2019a, doyle2019b}. The C/H axis extends down to $10^{-6}$ so that it reaches the carbon-poor compositions suggested by \citet{piaulet2024jwst}. A run ends when the planet reaches 5.465~Gyr, the system age estimated by \citet{passegger2024compact} (with an uncertainty of $\pm4.058$~Gyr), when the surface pressure drops below 3~bar (``escape stop''), or if the planet would be disrupted at the Roche limit or by rotational breakup. Mantle solidification does not end a run, so runs continue as solid planets with an outgassed atmosphere. Because every surviving run ends at the central age, the end-state comparison does not sample the age uncertainty. Sect.~\ref{subsec:results-reflation} examines the evolution across the $1\sigma$ age interval instead.

Table~\ref{tab:obs} summarises the observed properties of GJ~9827~d and how each enters the analysis. We compare the end state of each run with two primary constraints. The first is the bulk density of GJ~9827~d, $\rho = 2.454 \pm 0.739$~g\,cm$^{-3}$ \citep{passegger2024compact}. The second is the lower limit of 0.3162 ($\log_{10} > -0.5$) on the atmospheric H$_2$O volume mixing ratio (VMR) from the JWST/NIRISS retrieval of \citet{piaulet2024jwst}. This is a one-sided 95\% ($2\sigma$) lower limit from their retrieval with no assumed background gas (``agnostic''); their retrieval with a H$_2$/He background gives only $>0.0138$. We adopt the stricter agnostic limit, as a hard threshold, so the set of runs consistent with observations is conservative with respect to water. We define a simulation run to be density-consistent if its final $\rho_\mathrm{obs}$ lies within the $1\sigma$ interval (1.695--3.183~g\,cm$^{-3}$), and consistent with observations (hereafter a consistent run) if it also meets the water limit. Only runs that reach the present age can be consistent with observations, since a run that lost its atmosphere earlier ends as a bare rock with no water signal.

As a third, weaker constraint we use the 95\% upper limits on the CH$_4$ VMR from the same retrievals, $\log_{10} X_\mathrm{CH_4} < -3.60$ (agnostic background; $X_\mathrm{CH_4} < 2.5\times10^{-4}$) and $< -3.71$ (H$_2$/He background; $< 1.9\times10^{-4}$) \citep{piaulet2024jwst}. These limits rest mainly on the non-detection of the 2.3~$\mu$m CH$_4$ band and are described by the authors as tentative, so we apply them only after the density and water cuts and discuss their implications separately (Sect.~\ref{subsec:disc-methane}).

\section{Results}
\label{sec:results}

\begin{table*}
\caption{End states by mantle redox state.}
\label{tab:redox}
\centering
\begin{tabular}{r c c c c c c c c c c c}
\hline\hline
$\Delta$IW & $N_\mathrm{age}$ & $N_\mathrm{esc}$ & $\tilde{t}_\mathrm{esc}$ (Gyr) & H$_2$O & H$_2$ & CH$_4$ & CO$_2$ & $\tilde{\rho}_\mathrm{obs}$ (g\,cm$^{-3}$) & $N_{\Phi>0}$ & $N_\mathrm{cons}$ & $N_\mathrm{cons}^\mathrm{CH_4}$ \\
\hline
$-4$ & 3 & 101 & 0.0016 & 0.01 & 0.79 & 0.20 & 0.00 & 1.47 & 0 & 0 & 0 \\
$-3$ & 3 & 100 & 0.0098 & 0.02 & 0.78 & 0.20 & 0.00 & 1.44 & 0 & 0 & 0 \\
$-2$ & 6 & 99 & 0.11 & 0.06 & 0.75 & 0.18 & 0.00 & 1.72 & 0 & 0 & 0 \\
$-1$ & 42 & 63 & 0.20 & 0.23 & 0.74 & 0.03 & 0.00 & 1.74 & 18 & 0 & 0 \\
$0$ & 63 & 42 & 0.68 & 0.49 & 0.48 & 0.02 & 0.00 & 2.86 & 44 & 59 & 33 \\
$+1$ & 67 & 38 & 0.91 & 0.74 & 0.23 & 0.01 & 0.01 & 3.50 & 54 & 9 & 5 \\
$+2$ & 84 & 21 & 0.65 & 0.89 & 0.09 & 0.00 & 0.01 & 3.75 & 77 & 0 & 0 \\
$+3$ & 84 & 21 & 0.79 & 0.95 & 0.03 & 0.00 & 0.02 & 3.85 & 77 & 0 & 0 \\
$+4$ & 84 & 21 & 0.83 & 0.97 & 0.01 & 0.00 & 0.02 & 3.88 & 75 & 0 & 0 \\
\hline
\end{tabular}
\tablefoot{Each $\Delta$IW value has 105 cases. $N_\mathrm{age}$: runs reaching 5.465~Gyr; $N_\mathrm{esc}$: runs ending on the escape stop; $\tilde{t}_\mathrm{esc}$: median end time of the latter. The four runs missing at $\Delta$IW $= -4$ and $-3$ failed numerically (Sect.~\ref{subsec:results-overview}). The volume mixing ratios are means over the runs reaching 5.465~Gyr, and $\tilde{\rho}_\mathrm{obs}$ is their median final bulk density. $N_{\Phi>0}$: of these, runs with a partially molten mantle at 5.465~Gyr. $N_\mathrm{cons}$: runs consistent with observations, i.e. meeting the density and H$_2$O constraints; $N_\mathrm{cons}^\mathrm{CH_4}$: of these, runs that also meet the CH$_4$ upper limit ($X_\mathrm{CH_4} < 2.5\times10^{-4}$).}
\end{table*}

\subsection{Overview of the grid}
\label{subsec:results-overview}

Of the 945 cases, 942 completed: 436 reached the present age and 506 ended on the escape stop. Three cases stopped because the outgassing solver found no equilibrium solution. All three lie in one corner of the grid (15\,905~ppmw H, C/H~$=1$, $\Delta$IW~$=-4$ or $-3$). The solver failures occurred between 1.7 and 2.4~Gyr, at 12--21~bar over a solid mantle, shortly before these runs would probably have reached the escape stop. We exclude them from further analysis, which does not affect the set of runs consistent with observations. 

The outcome is set mainly by the mantle redox state (Table~\ref{tab:redox}). At $f\rm{O}_2 \le \Delta \rm{IW}$-2, 299 of the 311 completed runs lose their atmosphere, 225 of them within the first $10^8$~yr. At $f\rm{O}_2 \ge \Delta \rm{IW}$+2, 252 of 315 retain it to 5.465~Gyr. The hydrogen inventory is the secondary important parameter: all 189 runs with 1000~ppmw H end on the escape stop, the last of them (at $\Delta \rm{IW}$+4) after 2.5~Gyr. The surviving atmospheres are massive. At 5.465~Gyr their surface pressures range from $\sim10^2$ to $6\times10^5$~bar, and most of the oxidised runs still have a partially molten mantle, with global melt fractions $\Phi_\mathrm{global}$ of up to 0.33.

\subsection{Evolution of the bulk density}
\label{subsec:results-density}

   \begin{figure*}
   \centering
   \includegraphics[width=0.85\hsize]{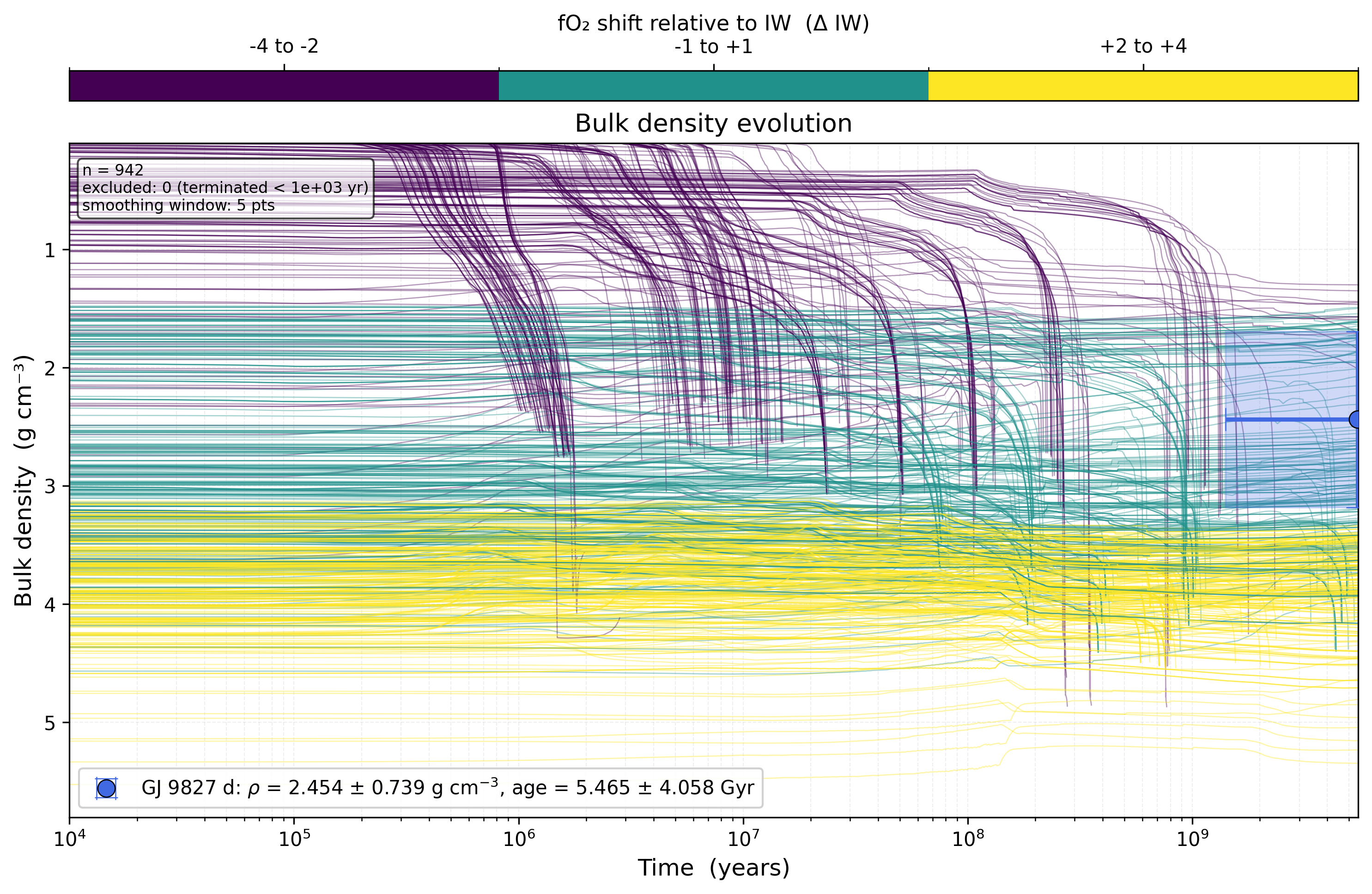}
      \caption{Bulk density $\rho_\mathrm{obs}$ against time for all 942 completed runs, coloured by mantle redox: $f\rm{O}_2$ from $\Delta$IW$-4$ to $\Delta$IW$-2$ (purple), $\Delta$IW$-1$ to $\Delta$IW$+1$ (teal) and $\Delta$IW$+2$ to $\Delta$IW$+4$ (yellow). The density axis is inverted, so that lower density (a more inflated planet) is up, and the earliest, most inflated states of the reduced runs lie above the top of the panel. Tracks are smoothed over five output points. Near-vertical drops mark runs whose atmosphere is stripped rapidly before the escape stop; the tracks reaching 5.465~Gyr are the runs that retain an atmosphere. The blue point, error bars and shaded box mark the measured density and age of GJ~9827~d with their $1\sigma$ uncertainties, $2.454\pm0.739$~g\,cm$^{-3}$ and $5.465\pm4.058$~Gyr \citep{passegger2024compact}.}
         \label{fig:rho_evo}
   \end{figure*}

Figure~\ref{fig:rho_evo} shows how the bulk density of every run evolves, and illustrates why redox is the first-order control. Most reduced runs (purple) start with extremely low densities: their H$_2$-rich, hot early atmospheres place the photosphere so high that the initial densities of the most extreme runs are below $\sim$1~g\,cm$^{-3}$. These runs then lose their envelopes through escape, and their densities rise steeply, by up to two orders of magnitude, before they reach the escape stop. 225 of the 299 stripped reduced runs do so before $10^8$~yr. The near-IW runs (teal) mostly start at $\sim$1.5--4.6~g\,cm$^{-3}$ and evolve more slowly. Of these, 143 are also stripped, 86 of them between $10^8$ and $10^9$~yr, while the remainder flatten out in or near the observed density range. The oxidised runs (yellow) start and remain dense, at 3.1--5.5~g\,cm$^{-3}$, because their H$_2$O-dominated envelopes are compact.

Relative to its own initial value, the density of a run that reaches the present age changes little: these runs end at 0.74--1.60 times their initial density, and 134 of 436 end below it. The runs that are stripped end far above their initial density. 
The present-day density of a surviving run is therefore set mostly by its starting volatile inventory and redox state. The exceptions to this rule are the runs that inflate late in their evolution, which we examine next.

\subsection{Reflation within the observed density and age range}
\label{subsec:results-reflation}

   \begin{figure*}
   \centering
   \includegraphics[width=0.85\hsize]{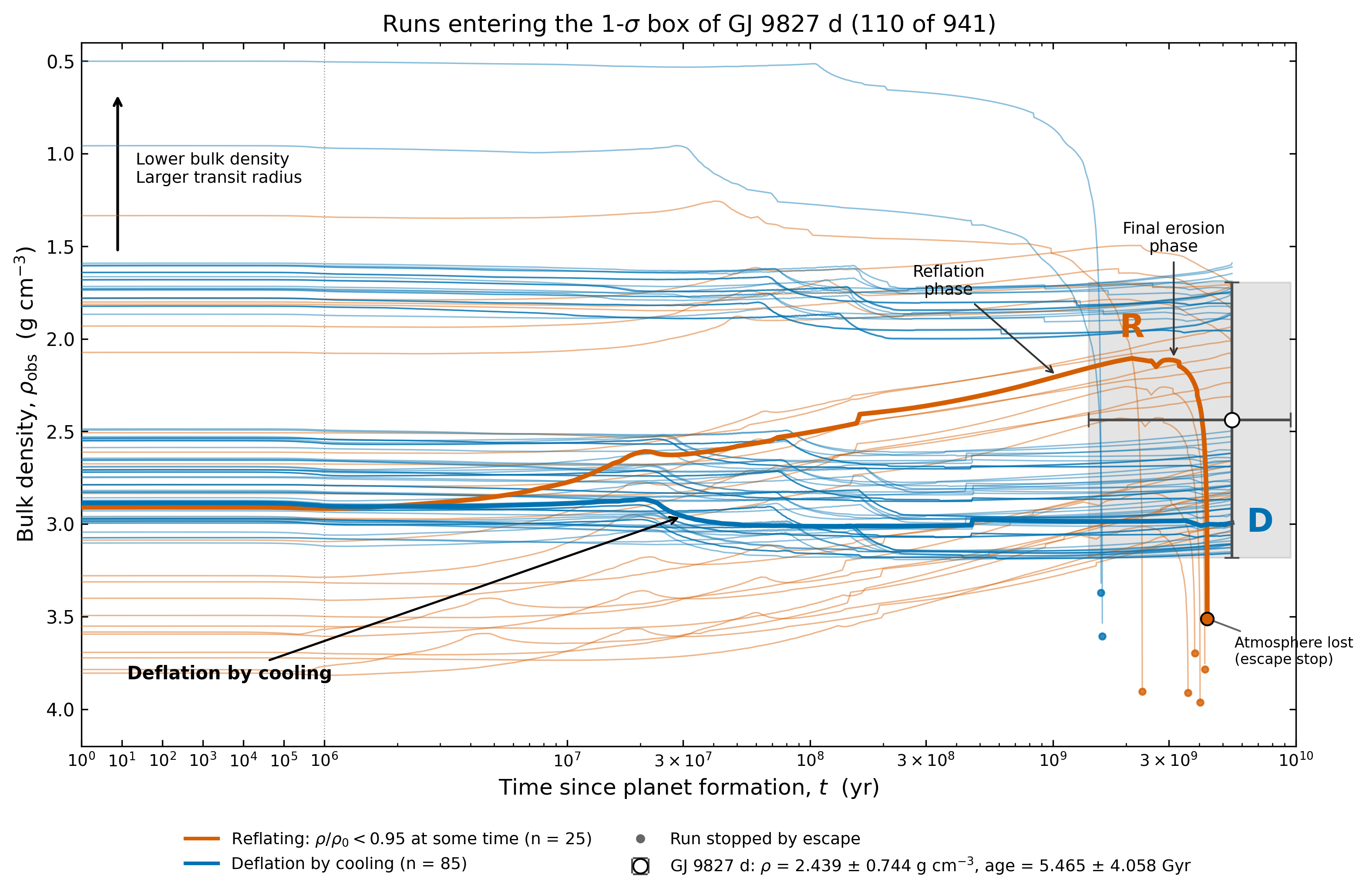}
      \caption{Bulk density against time for the 110 runs whose track enters, at any time, the $1\sigma$ box of the measured density and age of GJ~9827~d (grey box, point and error bars; $2.454\pm0.739$~g\,cm$^{-3}$ and $5.465\pm4.058$~Gyr). The first $10^6$~yr are compressed into the left fifth of the time axis (dotted line), and the density axis is inverted. Orange: reflating runs, whose density falls more than 5\% below its own initial value at some time (25 runs). Blue: the other 85 runs, which deflate as they cool. Dots mark runs that end on the escape stop. The thick lines are two examples: R (case 261; $\Delta$IW~$=-1$, 6325~ppmw H, C/H~$=1$, core mass fraction 0.215) reflates by 28\% before it loses its atmosphere, and D (case 269; $\Delta$IW~$=0$, 6325~ppmw H, C/H~$=10^{-2}$, core mass fraction 0.325) is the deflating run with the median final-to-initial density ratio. No run extends beyond 5.465~Gyr, so the part of the box at later ages is not sampled.}
         \label{fig:reflation}
   \end{figure*}

Fig.~\ref{fig:reflation} shows the 110 runs whose density passes through the $1\sigma$ box of the measured density and age (1.4--9.5~Gyr and 1.695--3.183~g\,cm$^{-3}$) at some time. Following \citet{cesario2026}, we separate them by the shape of their density evolution. We call a run reflating if its density falls more than 5\% below its own initial value at any time. The threshold excludes small transient dips.

Eighty-five runs do not reflate. Their density stays within 1\% of its initial value for the first $\sim$1~Myr and then rises by a few per cent, which we attribute to cooling of the mantle and atmosphere, and the 83 of them that reach the present age end at 0.95--1.10 times their initial density (median 1.04). We refer to this behaviour as deflation by cooling. Run D in Fig.~\ref{fig:reflation} is a typical example.

The other 25 runs reflate, and they form a well-defined group. All of them have mantle $f\rm{O}_2$ between $\Delta$IW$-3$ and $\Delta$IW$\pm$0, and 24 of the 25 have the highest carbon inventory, C/H~$=1$; the remaining one has C/H~$=0.1$. Their density falls to 0.72--0.94 times its initial value. Nineteen of them reach the present age, 15 of these while still at or near their minimum density, so that they are observed in a reflated state. The other six reflate, then compress again as their atmosphere is lost, and end on the escape stop at 2.3--4.3~Gyr. Run R shows the full sequence: its density falls from 2.91 to 2.11~g\,cm$^{-3}$ by 2.1~Gyr and then rises to 3.51~g\,cm$^{-3}$ when the atmosphere is lost at 4.3~Gyr.

In the 24 carbon-rich reflating runs, the drop in density coincides with a change of the atmosphere from a carbon-bearing to a hydrogen-dominated composition. Between the density maximum that precedes reflation and the density minimum, the CO mixing ratio falls from 5--61\% to at most 5.4\%, while H$_2$ and CH$_4$ rise. The mean molecular weight falls by a median factor of 0.64 (range 0.40--0.96), from 6.4--25~g\,mol$^{-1}$ to 5.8--11.7~g\,mol$^{-1}$. Over the same interval the surface cools from 3850--4410~K to 1900--2400~K in 22 of the 24 runs, and the melt fraction falls from $\ge 0.97$ to at most 0.58. In run R, for instance, the CO fraction falls from 35\% to 2.5\%, H$_2$ rises from 42\% to 68\% and CH$_4$ from 1\% to 10\%, the mean molecular weight falls from 15.2 to 7.3~g\,mol$^{-1}$, the surface pressure falls from $2.9\times10^4$ to $5.3\times10^3$~bar, and the melt fraction falls from 0.99 to 0.20. We discuss the mechanism in Sect.~\ref{subsec:disc-reflation}.

Eight of the reflating runs are also consistent with observations (Sect.~\ref{subsec:results-filter}). All have $f\rm{O}_2 = \Delta$IW$\pm0$ and C/H~$=1$, and their present-day densities, 2.66--3.04~g\,cm$^{-3}$, are 0.79--0.81 times their initial densities of 3.31--3.81~g\,cm$^{-3}$. Without reflation, these runs would lie at or above the upper edge of the density interval. The remaining 11 reflating runs that reach the present age, all at $f\rm{O}_2 = \Delta$IW-2 or $\Delta$IW-1, are consistent with the density but not with the water limit.

\subsection{Observational filtering}
\label{subsec:results-filter}

   \begin{figure}
   \centering
   \includegraphics[width=\hsize]{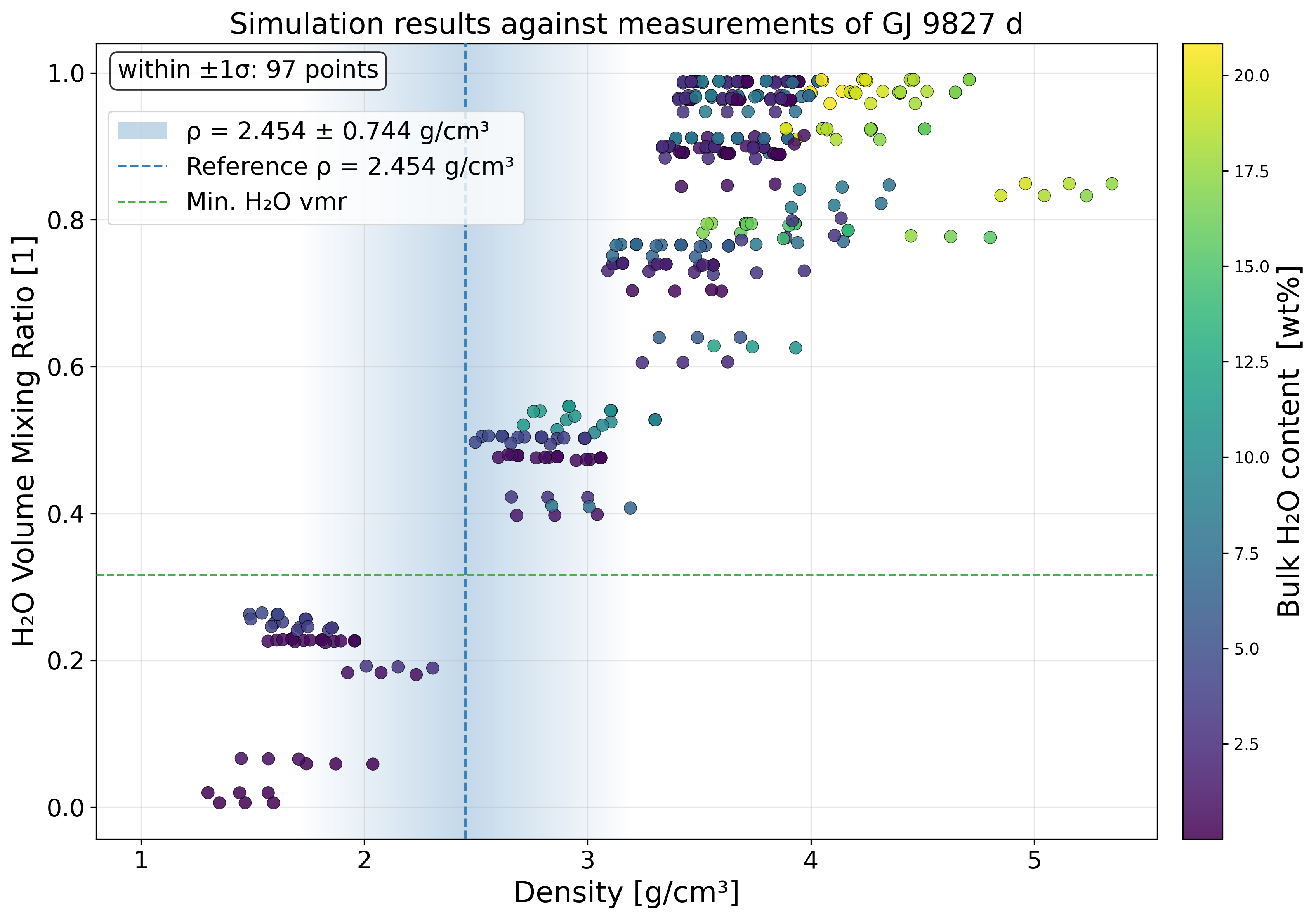}
      \caption{End states of the 436 runs that reach 5.465~Gyr: atmospheric H$_2$O volume mixing ratio against bulk density, coloured by bulk water content $100\,M_\mathrm{H_2O}/M_\mathrm{p}$ (wt\%, counting atmosphere and mantle). The dashed blue line and shaded band mark the measured density and its $1\sigma$ interval, $2.454\pm0.739$~g\,cm$^{-3}$ \citep{passegger2024compact}; the dashed green line marks the minimum H$_2$O VMR of 0.3162 \citep{piaulet2024jwst}. The 98 runs within the density band include the 68 that also meet the water limit (upper part of the band). Runs that differ only in C/H at C/H~$\le 10^{-3}$ end at almost identical positions and overlap.}
         \label{fig:all}
   \end{figure}

Figure~\ref{fig:all} reduces the end state of every run that reaches the present age to the two primary observables. The runs form distinct clumps, each corresponding largely to one $\Delta$IW value, because redox fixes both the atmospheric composition and, through the mean molecular weight, the scale height and hence the radius. Of the 436 runs, 98 lie within the $1\sigma$ density interval and 382 meet the water limit, but only 68 do both. The 30 runs consistent with the density only lie at 1.70--2.31~g\,cm$^{-3}$ with H$_2$O VMRs of 0.06--0.26; all are at $\Delta$IW~$=-1$ or $-2$ and have $\ge 15\,905$~ppmw H, so their envelopes are light enough to be inflated but too poor in water. The oxidised runs meet the water limit easily, with H$_2$O VMRs of 0.73--0.99, but their compact steam envelopes make them too dense (3.3--5.3~g\,cm$^{-3}$). No run that ends on the escape stop meets both constraints.

The 68 runs consistent with observations have bulk densities of 2.50--3.16~g\,cm$^{-3}$, on the dense side of the measured value, and H$_2$O VMRs of 0.40--0.77. Their bulk water contents span 0.6--11.6~wt\% (median 3.9~wt\%), and their atmospheres hold 0.4--13\% of the planet's mass (median 3\%). At 5.465~Gyr their surface pressures are $7.5\times10^3$--$2.4\times10^5$~bar and their surface temperatures 1950--2680~K, so in the model the observed atmosphere lies on top of a deep, hot envelope.

To rank the consistent runs, we weighted each by a Gaussian likelihood in density, $\ln\mathcal{L} = -\tfrac{1}{2}[(\rho_\mathrm{obs}-2.454)/0.739]^2$, with the water limit applied as a step function. The highest-ranked runs lie closest to the measured density, at 2.50--2.62~g\,cm$^{-3}$. The top six all have $\Delta$IW~$=0$ and the smallest core fraction (0.215), and five of them 15\,905~ppmw H. Their mean atmosphere is 50\% H$_2$O and 49\% H$_2$, with 0.4\% CH$_4$. The ranking thus selects the same family of solutions as the simple cut, favouring smaller cores and intermediate hydrogen inventories within it. With only one density measurement, though, the likelihood differences across the band are modest ($\Delta\ln\mathcal{L} < 0.5$), and we do not regard this ranking as a constraint on the core fraction.

\subsection{Parameter dependence}
\label{subsec:results-params}

   \begin{figure*}
   \centering
   \includegraphics[width=0.9\hsize]{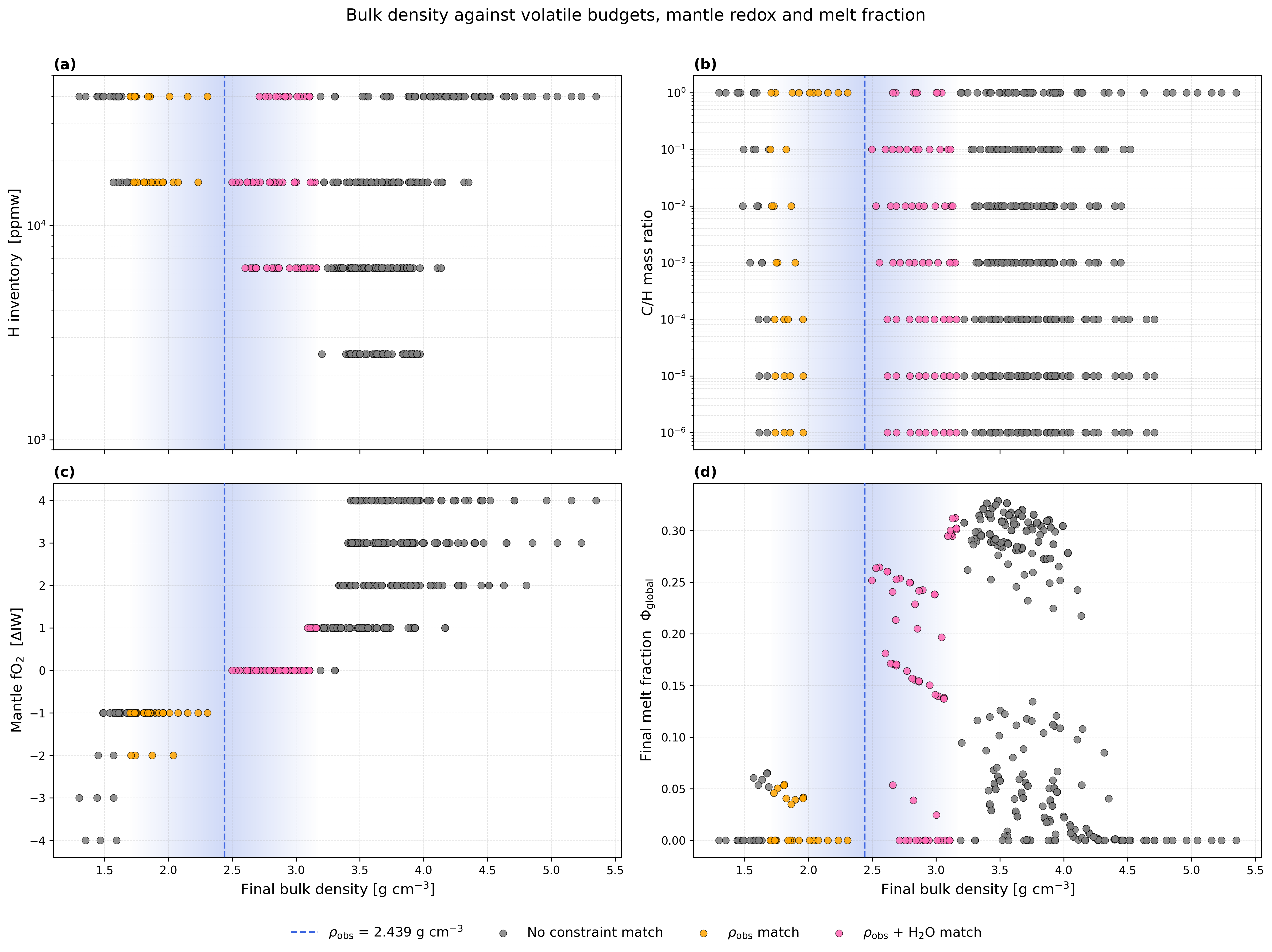}
      \caption{Final bulk density of the 436 runs that reach 5.465~Gyr against (a) the initial hydrogen inventory, (b) the C/H mass ratio, (c) the mantle redox state $\Delta$IW, and (d) the final global melt fraction $\Phi_\mathrm{global}$. Pink: runs meeting both the density and the H$_2$O constraint; orange: density only; grey: neither. The dashed line and shading mark the measured density, $2.454\pm0.739$~g\,cm$^{-3}$ \citep{passegger2024compact}. Panels (a) and (c) show that the runs consistent with observations require $\ge 6325$~ppmw H and $\Delta$IW~$=0$ or $+1$. Panel (b) shows consistent runs at every C/H value from $10^{-6}$ to 1, so density and water alone do not constrain the carbon inventory. Panel (d) shows that most consistent runs retain a partially molten mantle.}
         \label{fig:fourpanel}
   \end{figure*}

Figure~\ref{fig:fourpanel} resolves the set of consistent runs by grid axis. Redox (panel c) separates the runs most cleanly. All consistent runs have $f\rm{O}_2 = \Delta$IW$\pm0$ or $\Delta$IW+1. At $\Delta$IW$\pm0$, 59 of the 63 runs that reach the present age are consistent; the four exceptions all have 40\,000~ppmw H and the largest core fraction, and lie just above the upper edge of the density band (3.19--3.30~g\,cm$^{-3}$). At $\Delta$IW+1 only 9 of 67 are consistent, all with the smallest core fraction and at 3.09--3.16~g\,cm$^{-3}$, within 0.1~g\,cm$^{-3}$ of the upper edge. This makes the consistent runs at $\Delta$IW+1 sensitive to the density bias discussed in Sect.~\ref{subsec:disc-limits}.

The hydrogen inventory (panel a) sets a lower bound. No run with 2515~ppmw H or less is consistent with the observations: at 1000~ppmw all runs lose their atmosphere, and at 2515~ppmw the survivors, which are all at $f\rm{O}_2\ge\Delta$IW+1, are too dense. 
Above the bound, the consistent runs are spread over 6325, 15\,905 and 40\,000~ppmw (27, 24 and 17 runs); within the near-IW runs, the final density depends only weakly on the hydrogen inventory.

Neither the carbon inventory nor the core mass fraction is constrained by density and water. Consistent runs occur at every C/H value (panel b), with 9 runs at each of C/H~$=10^{-6}$, $10^{-5}$ and $10^{-4}$, 11 at each of $10^{-3}$, $10^{-2}$ and $10^{-1}$, and 8 at C/H~$=1$. Below C/H~$\approx 10^{-3}$ carbon is a trace constituent of the atmosphere, and runs that differ only in C/H end with indistinguishable densities and water fractions. Even at C/H~$=0.1$ carbon enters the atmospheres of the consistent runs mainly as CH$_4$, at no more than 1.7\%, which changes the mean molecular weight too little to move a planet across the density bound. Only at C/H~$=1$, where the atmospheres of the consistent runs hold 12--19\% CH$_4$, does carbon change the outcome noticeably: the H$_2$O fraction drops to 0.40--0.42, and all eight consistent runs at this C/H are reflating (Sect.~\ref{subsec:results-reflation}). All three core fractions produce consistent runs (30, 21 and 17 runs at 0.215, 0.27 and 0.325), with a mild preference for the smallest core, which gives a larger, less dense interior.

Panel (d) shows the final melt fraction. Fifty-three of the 68 consistent runs still have a partially molten mantle at 5.465~Gyr, with $\Phi_\mathrm{global}$ of up to 0.31. The other 15 are fully solid; all of these have $f\rm{O}_2=\Delta$IW$\pm0$ and 40\,000~ppmw H, and their thick envelopes come in part from the release of the whole dissolved volatile inventory at solidification (Sect.~\ref{subsec:disc-limits}).

\subsection{Mantle redox and atmospheric composition}
\label{subsec:results-redox}

   \begin{figure}
   \centering
   \includegraphics[width=\hsize]{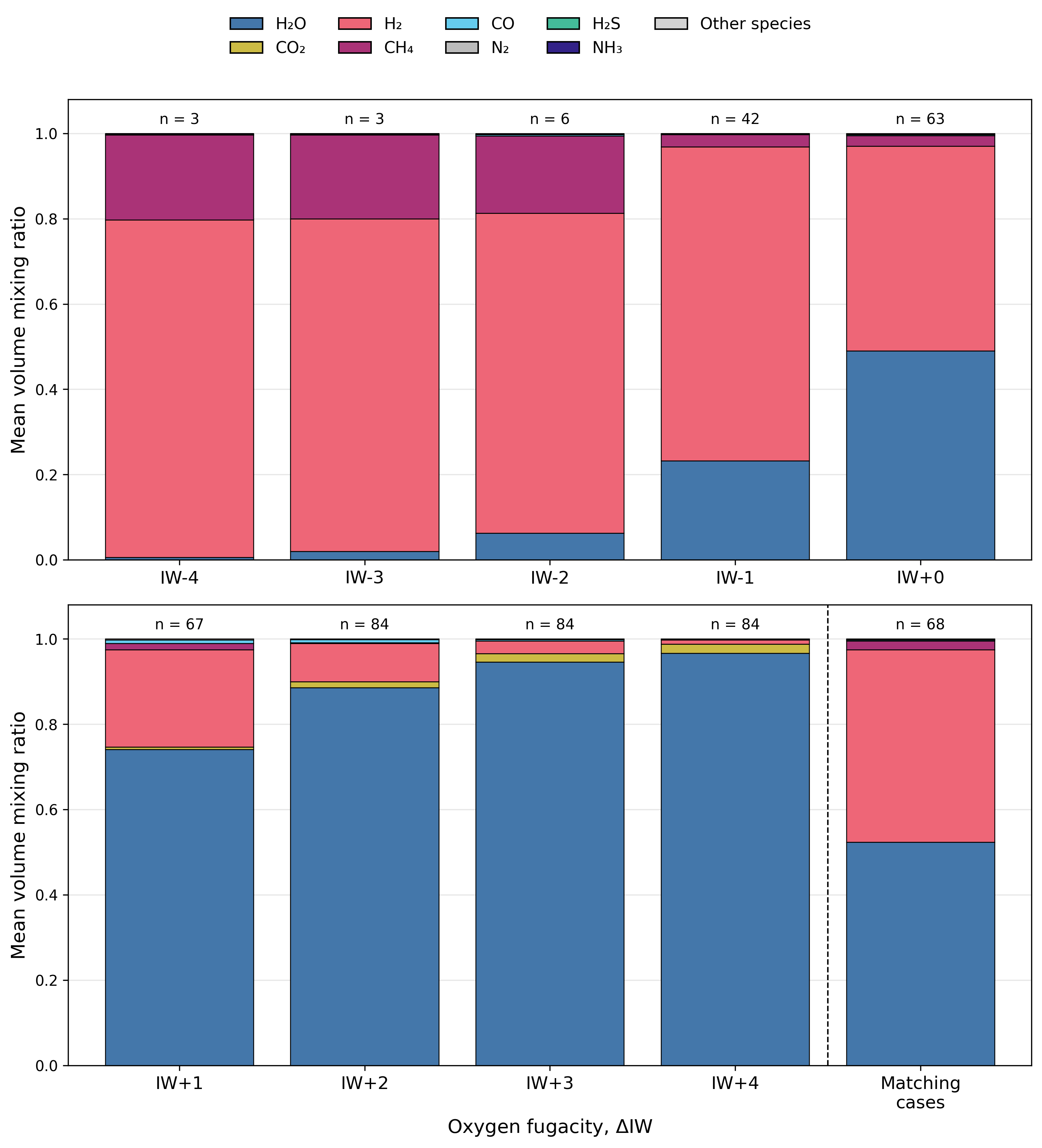}
      \caption{Mean atmospheric composition (volume mixing ratio) at 5.465~Gyr of the runs that reach the present age, grouped by mantle redox state ($\Delta$IW from $-4$ to 0, top; $+1$ to $+4$, bottom). The right-hand column of the bottom panel shows the mean composition of the 68 runs consistent with observations, which have $\Delta$IW~$=0$ (59 runs) or $+1$ (9 runs). Each column shows its five most abundant species and groups the remainder as ``other''. Species below $\sim$1\% are too thin to be visible; the mean atmosphere of the consistent runs is effectively a three-component H$_2$O--H$_2$--CH$_4$ mixture, whose CH$_4$ comes almost entirely from the runs with C/H~$\ge 0.1$. $n$ is the number of runs averaged in each column.}
         \label{fig:fO2_all}
   \end{figure}

Figure~\ref{fig:fO2_all} shows how the mean atmospheric composition of the surviving runs changes with the mantle's redox state. The trend is monotonic, as expected for magma-ocean-derived atmospheres, whose mean molecular weight is set mainly by the oxygen fugacity \citep{sossi2020, gaillard2021, suer2023, sossi2023, nicholls2024a, tian2024, heng2025}, but the reduced end differs from the conventional picture of CO-rich reduced atmospheres. At $f\rm{O}_2=\Delta$IW-4 to $\Delta$IW-2 the atmospheres are about 75--80\% H$_2$ and 18--20\% CH$_4$, with H$_2$O at only 1--6\%. The H$_2$O fraction then rises steadily with oxygen fugacity: 23\% at $\Delta$IW-1, 49\% at 0, 74\% at $\Delta$IW+1, 89\% at $\Delta$IW+2, 95\% at $\Delta$IW+3 and 97\% at $\Delta$IW+4. CH$_4$ declines in step, and CO$_2$ reaches only a few per cent at the oxidised end (2\% at $\Delta$IW+4). The mean CO fraction stays below 1\% at every oxidation state.

Three effects in the coupled outgassing calculation produce this trend. First, higher oxygen fugacity shifts the gas-phase equilibria towards oxidised species (H$_2\rightarrow$H$_2$O, CH$_4\rightarrow$CO$_2$). Second, CH$_4$ is favoured at the low oxygen fugacity, high pressure and abundant hydrogen of the reduced runs \citep{tian2024, heng2025}, so the carbon of these runs is held mostly as CH$_4$ rather than CO at the present age. Third, the survivors at low $f\rm{O}_2$ are a strongly selected subset: only 12 of 311 reduced runs reach the present age, all with large hydrogen inventories and high C/H, so their mean composition reflects the H- and C-rich end of the grid.

The 68 consistent runs lie at the transition. Their mean atmosphere is 52\% H$_2$O, 45\% H$_2$ and 2\% CH$_4$, with CO, CO$_2$, NH$_3$ and H$_2$S below 0.3\% each; individual runs range from 40\% to 77\% H$_2$O. In this sense GJ~9827~d is a ``steam world'' only in part: water is the most abundant species, but molecular hydrogen is almost as abundant. The corresponding mean molecular weights are 9.6--14.6~g\,mol$^{-1}$.

The CH$_4$ fraction in the atmosphere of a consistent run is set almost entirely by the initial carbon inventory. In these runs most of the atmospheric carbon is held as CH$_4$ (median 91\%), and the atmospheric C/H mass ratio is within about 6\% of the initial one, so the CH$_4$ VMR scales nearly linearly with C/H: it is about 0.16 times the C/H mass ratio at $\Delta$IW$\pm0$ and about 0.03--0.1 times it at $\Delta$IW$+1$. The consistent runs therefore span seven orders of magnitude in CH$_4$, from below $10^{-7}$ at C/H~$=10^{-6}$ to 0.19 at C/H~$=1$. All 38 consistent runs with C/H~$\le 10^{-3}$ have CH$_4$ VMRs below both JWST upper limits (Sect.~\ref{subsec:sims}), while all 30 with C/H~$\ge 10^{-2}$ exceed them, the least CH$_4$-rich of these by about 20\% ($3.0\times10^{-4}$). The 38 runs that meet all three constraints have $\Delta$IW$\pm0$ (33) or $\Delta$IW+1 (5), 6325--40\,000~ppmw H, densities of 2.56--3.16~g\,cm$^{-3}$ and H$_2$O VMRs of 0.47--0.77. 31 of them still have a partially molten mantle.

\section{Discussion}
\label{sec:discussion}

\subsection{Mass adopted for GJ~9827~d}
\label{subsec:disc-mass}

We adopted the mass from \citet{passegger2024compact}, $M_\mathrm{p} = 3.02^{+0.58}_{-0.57}\,M_\oplus$, rather than the values of \citet{rice2019masses} and \citet{kosiarek2020physical}. \citet{passegger2024compact} combined ESPRESSO radial velocities with a Gaussian-process model of stellar activity constrained jointly by the radial velocities and photometry, which treats the correlated activity noise of the host more completely than earlier analyses and yields smaller uncertainties on the planetary parameters. All runs start at this mass. Because the planet mass in the model then decreases (Sect.~\ref{subsec:disc-limits}), the consistent runs end at 2.66--2.96~$M_\oplus$, still within the $1\sigma$ interval of the measurement. Their transit radii, 1.70--1.85~$R_\oplus$, are correspondingly smaller than the measured 1.89~$R_\oplus$. A direct comparison with the radius would be a stricter test than the density alone.

\subsection{The interior implied by the runs consistent with observations}
\label{subsec:disc-interior}

The runs consistent with observations give a model-based picture of the interior of GJ~9827~d. In all of them the planet carries several wt\% of water (median 3.9~wt\%) together with an atmosphere of a few per cent of its mass. This is far below the $32\pm10$\% that \citet{piaulet2024jwst} inferred, and we attribute the difference to two fundamental differences in model construction. First, their estimate comes from a static structure model, which fits the present-day mass and radius with an interior in its final state (a fully solidified mantle). Our framework instead evolves the planet forward in time, self-consistently coupling interior thermal evolution, magma-ocean solidification, volatile exchange between melt and atmosphere, and atmospheric structure up to the observed age of 5.5~Gyr. Second, \citet{piaulet2024jwst} place all of the water in an envelope of pure H$_2$O vapour or supercritical fluid on top of a rock--iron core. Such an envelope has a high mean molecular weight and is therefore compact per unit mass, so a large water inventory is required to reproduce the observed radius. In our runs the envelope is about half H$_2$ by volume and thus more extended per unit mass, and a substantial part of the water is dissolved in the mantle rather than residing in the envelope. The inferred bulk water content therefore drops by roughly an order of magnitude. It is, on the other hand, above the $<1.5$~wt\% that \citet{werlen2025sub} found in the picture of global chemical equilibration of sub-Neptunes, a calculation that assumes constant hydrogen and oxygen sinks in the metal phase of the interior \citep{Lichtenberg2021ApJL} that our model does not account for.

Because volatiles dissolve in the melt, a large fraction of the water need not be in the atmosphere: in the consistent runs that are still partially molten, the mantle holds up to 61\% of the bulk water (median 37\% over all 68). A partially molten mantle at 5.5~Gyr thus acts as a large, still exchanging volatile reservoir, a role that JWST observations and predictions also suggest for the magma oceans of the ultrahot super-Earth TOI-561~b \citep{teske2025,Postolec2026arXiv} and other lava worlds \citep{Sastre2026arXiv}. This follows from the thick H$_2$O--H$_2$ envelope, whose greenhouse effect keeps the surface at about 1950--2700~K, well above the solidus, in line with the persistence of magma oceans under steam-rich and hydrogen-bearing atmospheres on strongly irradiated planets \citep{hamano2015, lichtenberg2021, innes2023runaway, nicholls2024b, boer2025}. Coupled models of H$_2$-dominated sub-Neptunes likewise find that most known sub-Neptunes would host permanent magma oceans if they are gas dwarfs \citep{calder2026}, although that study explicitly excludes high-mean-molecular-weight planets such as GJ~9827~d.

Whether this melt survives in reality is less certain than the model suggests. The release of all remaining dissolved volatiles at solidification (Sect.~\ref{subsec:disc-limits}) and the absence of volcanic, rate-limited degassing and of solid-phase volatile storage all affect how much gas sits above a cooling mantle. In addition, the surface pressures of $10^4$--$10^5$~bar place the base of the envelope in a regime where the ideal-gas treatment and a sharp melt--gas interface both require large extrapolations in measured physical constants (e.g. partition coefficients of volatile species) \citep{innes2023runaway, benneke2024jwst, bower2025}. The melting curves of the mantle may themselves depend on oxygen fugacity, but this matters mainly for volatile-poor planets, whereas the mantles of volatile-rich planets stay far above the melting region \citep{sastre2026}.

The core mass fraction and hydrogen inventory trade off against each other only weakly in the consistent set: all three core fractions produce consistent runs at all three admissible H inventories. Thus bulk density and the water limit alone cannot fix the core mass fraction, in line with the degeneracy between core fraction and volatile budget found in evolutionary retrievals of sub-Neptunes \citep{nicholls2026retrieval}. Independent constraints, for instance from refractory abundances of the host star \citep[as used for the TOI-1064 planets by][]{wilson2022}, would be needed.

\subsection{Atmospheric composition and the JWST water signature}
\label{subsec:disc-atmos}

The mean composition of the consistent runs (52\% H$_2$O, 45\% H$_2$, 2\% CH$_4$) is neither a pure steam envelope nor a primordial hydrogen envelope. It meets the JWST/NIRISS lower limit on water by construction, but its mean molecular weight, 9.6--14.6~g\,mol$^{-1}$, is lower than the $18.02^{+0.20}_{-0.61}$~g\,mol$^{-1}$ (68\% interval) of the agnostic-background retrieval of \citet{piaulet2024jwst}, which corresponds to an almost pure water atmosphere. It is, however, consistent with the $9.84^{+4.51}_{-4.64}$~g\,mol$^{-1}$ obtained when the same data are fitted with a H$_2$/He background. The comparison therefore depends on the assumed background gas, and the data do not by themselves exclude the mixed H$_2$O--H$_2$ envelopes found here. Only the agnostic retrieval is in tension with them. Hydrocarbon hazes, an alternative explanation for muted sub-Neptune spectra, are expected to be suppressed at low C/O ratios, and the low spectral amplitude of GJ~9827~d is more naturally explained by a high mean molecular weight \citep{yang2026soot}. Two caveats apply to this comparison. First, the retrieval constrains the composition in the pressure range probed by transmission spectroscopy, whereas our model assumes a well-mixed atmosphere from the surface up. Condensation, photochemistry, or compositional gradients in such a deep envelope could raise the water fraction near the photosphere above its bulk value, even though recent work indicates this to be only of minor effect \citep{Panagiotou2026arXiv}. Second, the model's density bias (Sect.~\ref{subsec:disc-limits}) means that more H$_2$-poor atmospheres than those of the consistent set may in reality be compatible with the measured density. Both effects would push the solution towards higher $f\rm{O}_2$ and higher water fractions. A tighter test would compare synthetic transmission spectra of the consistent runs directly with the NIRISS data rather than relying on VMR thresholds.

The H$_2$ fraction of the consistent runs also matters for escape. The non-detections of extended H and He exospheres \citep{kasper2020nondetection, carleo2021multiwavelength, krishnamurthy2023absence} constrain rapidly escaping, hydrogen-dominated envelopes. In our model the consistent runs lose 6--45\% of their initial volatile mass over 5.5~Gyr through energy-limited escape, and the remaining envelopes are $\sim$45\% H$_2$ by volume but only $\sim$10\% by mass. Whether such an envelope, replenished from the mantle, would produce a detectable Ly$\alpha$ or He\,\textsc{i} signal at the present escape rate requires a dedicated calculation of the upper atmosphere and outflow, which we leave to future work. In water-rich atmospheres, hydrogen lost through water photolysis may escape mainly in ionised form \citep{Luger2015AsBio,Wordsworth2018AJ,Attia2026arXiv}, which these observations do not constrain \citep{piaulet2024jwst}.

\subsection{Methane as a constraint on the carbon inventory}
\label{subsec:disc-methane}

Bulk density and atmospheric water constrain the redox state and the hydrogen inventory of GJ~9827~d but say almost nothing about its carbon: consistent runs exist at every C/H from $10^{-6}$ to 1 (Sect.~\ref{subsec:results-params}). The CH$_4$ upper limit of \citet{piaulet2024jwst} adds information precisely along this otherwise unconstrained axis. Because the carbon in the atmospheres of the consistent runs is held mainly as CH$_4$, as expected at sub-Neptune surface pressures \citep{heng2025}, and because escape does not fractionate carbon from hydrogen in our model, the CH$_4$ VMR at 5.465~Gyr tracks the initial C/H ratio closely. The limit of $X_\mathrm{CH_4} < 2.5\times10^{-4}$ then translates into C/H~$\lesssim 10^{-3}$ for the consistent runs on our grid: all runs at C/H~$\le 10^{-3}$ pass, and all at C/H~$\ge 10^{-2}$ fail. Interpolating the near-linear scaling between CH$_4$ and C/H places the threshold at C/H~$\approx 1.6\times10^{-3}$ for $\Delta$IW$\pm0$ and $\approx 8\times10^{-3}$ for $\Delta$IW$+1$, where less of the carbon is in CH$_4$. The stricter H$_2$/He-background limit ($1.9\times10^{-4}$) selects the same runs. The methane limit leaves the redox and hydrogen constraints intact, since the 38 remaining runs span the same $f\rm{O}_2$ values, hydrogen inventories and core fractions as the full consistent set. It does, however, remove all eight reflated consistent runs, whose atmospheres hold 12--19\% CH$_4$ (Sect.~\ref{subsec:disc-reflation}). The atmospheric C/H mass ratio retrieved by \citet{piaulet2024jwst}, $\log_{10}(\mathrm{C/H}) = -2.8\pm2.0$ (agnostic background) or $-2.9\pm2.0$ (H$_2$/He background), agrees with the atmospheres of our consistent runs for C/H between $\sim10^{-5}$ and $\sim10^{-1}$ at $1\sigma$, so in this case the CH$_4$ limit is the more informative of the two carbon constraints.

Taken at face value, this would mean that GJ~9827~d formed, or ended up, with a mantle--atmosphere system poor in carbon relative to hydrogen, by at least two to three orders of magnitude below the upper end of our grid. Several caveats make this a weaker constraint than density and water, which is why we treat it as a third, minor one. First, the limit itself is tentative: it rests on the non-detection of a single CH$_4$ band, and the chemically consistent retrievals of \citet{piaulet2024jwst} only marginally ($\sim1.9\sigma$) favour CH$_4$ depletion relative to chemical equilibrium. Second, our atmospheres are well-mixed, with a composition set by equilibrium at the surface, at $\sim$2000~K and $10^4$--$10^5$~bar. In the much cooler photosphere probed by NIRISS, thermochemistry and quenching would redistribute carbon among CH$_4$, CO and CO$_2$ \citep{heng2025}. In temperate sub-Neptunes this partitioning follows the deep H$_2$O/H$_2$ ratio, with CH$_4$ and CO$_2$ of comparable abundance near H$_2$O/H$_2\approx1$ \citep{yang2024}, close to the ratio in our consistent runs. Photochemistry could also deplete CH$_4$ in the upper atmosphere and leave CO and CO$_2$ as the dominant observable carbon carriers \citep{werlen2025co}, although for the sub-Neptune TOI-270~d \citet{nixon2025} found the H$_2$O and CH$_4$ abundances largely unaffected by vertical mixing and photochemistry. We have not modelled these effects, so the mapping from bulk to photospheric CH$_4$ is uncertain in both directions. Third, carbon could be stored in reservoirs our model does not include, for instance in the metal phase during core formation or as graphite \citep{werlen2025co, bower2025}, so that a low atmospheric C/H need not imply a low bulk C/H. Finally, \citet{piaulet2024jwst} suggested that the absence of CH$_4$ and NH$_3$ could instead reflect an oxidised, H$_2$O--O$_2$ atmosphere, produced when hydrogen escape overwhelms the oxygen sinks of the mantle; our model, with a buffered $f$O$_2$ and non-fractionating escape, cannot produce such a state. A methane constraint of this kind would become decisive if the planned NIRSpec observations \citep[GO~4098;][]{piaulet2024jwst} confirm or tighten the limit and if it is compared with synthetic spectra of the consistent runs that include vertical chemistry.

\subsection{Reflation and the present-day density of GJ~9827~d}
\label{subsec:disc-reflation}

\citet{cesario2026} showed that irradiated super-Earths with reduced mantles near the IW buffer can inflate transiently: escape removes an initially high mmw atmosphere, the falling surface pressure releases dissolved water from the magma ocean, and at the prevailing low oxygen fugacity much of it is outgassed as H$_2$. The lighter gas lowers the mean molecular weight and raises the scale height, and in their models the bulk density drops by up to $\sim$60\% before the atmosphere is finally lost. Oxidised mantles, whose heavy CO$_2$- and SO$_2$-rich atmospheres buffer the mean molecular weight, deflate monotonically instead. Their reflation is favoured by mantles at $f\rm{O}_2 \approx \Delta$IW$\pm0$, by initial hydrogen inventories above about five Earth oceans ($\approx$100 ppmw), by high irradiation and by intermediate escape efficiencies.

Our runs for GJ~9827~d, a more massive planet than the fiducial case of \cite{cesario2026}, show the same qualitative behaviour. Among the runs that pass through the measured density and age, reflation occurs only at $f\rm{O}_2 \le \Delta$IW$\pm0$, its onset coincides with the loss of CO from the atmosphere and its replacement by H$_2$, and the mean molecular weight falls by up to 60\% while the density falls by up to 28\%. Six of the reflating runs then lose their atmosphere entirely, which reproduces the sequence of reflation, final erosion and complete loss described by \citet{cesario2026}. There are also differences. Reflation in our grid requires a high carbon inventory: 24 of the 25 reflating runs have C/H~$=1$, which is what makes the early atmosphere CO-rich in the first place, and the change of composition coincides with strong surface cooling (from $\sim$4000 to $\sim$2000~K) and with crystallisation of much of the mantle. Part of the shift from CO to H$_2$ and CH$_4$ is therefore probably thermochemical, as CH$_4$ becomes stable at lower temperature, and part may come from the release of dissolved volatiles as the melt fraction drops, in addition to the pressure decrease caused by escape.

For the interpretation of GJ~9827~d, reflation matters because it adds a degeneracy to the density constraint. Eight of the 68 consistent runs reach the measured density only because they are reflated at the present age: they started at 3.3--3.8~g\,cm$^{-3}$, above the $1\sigma$ interval, and their density has since fallen by about 20\%. A low bulk density can thus indicate either a planet on a monotonic cooling track or one observed during a transient reflation episode, and the density alone cannot distinguish the two. In our grid the atmospheric composition breaks this degeneracy, as evolutionary considerations have been suggested to do for GJ~9827~d and similar planets \citep{lichtenberg2023}: the reflated cases consistent with the observations hold 12--19\% CH$_4$ and 40--42\% H$_2$O, the lowest water fractions of the consistent set, and the CH$_4$ limit of \citet{piaulet2024jwst} excludes them. If future observations confirm the non-detection of CH$_4$ and that limit holds, GJ~9827~d is more likely on the deflation branch of Fig.~\ref{fig:reflation} than in a reflated state. This conclusion inherits the caveats of the methane limit (Sect.~\ref{subsec:disc-methane}), and it applies only to the carbon-rich, CO-driven reflation found here. A more precise age would also help, since our $1\sigma$ age interval spans 1.4--9.5~Gyr and covers both reflating and deflating phases. PLATO aims to determine stellar ages to about 10\% for Sun-like stars \citep{rauer2025}, although whether this precision can be reached for a K dwarf such as GJ~9827 is not established.

\subsection{Redox state and the formation context of GJ~9827~d}
\label{subsec:disc-redox}

The consistent runs require a mantle within about one log unit of the IW buffer. Strongly reduced mantles ($f\rm{O}_2 \le \Delta IW$$\pm0$) produce H$_2$-dominated envelopes that are either stripped within $\sim$0.1~Gyr or, if they survive, too inflated and too poor in water. Oxidised mantles ($f\rm{O}_2 \ge \Delta IW$+2) produce compact steam envelopes that are too dense. Coupled evolution models of other well-characterised planets point to a range of mantle redox states. For the molten super-Earth L~98-59~d, \citet{nicholls2026} required $f$O$_2$ between $\Delta$IW$-4$ and $\Delta$IW$-1$ to reproduce its density and mean molecular weight. For the lava world TOI-561~b, \citet{Postolec2026arXiv} favoured an oxidised mantle ($f$O$_2\gtrsim\Delta$IW+4) beneath a CO$_2$- or SO$_2$-rich atmosphere; and for the sub-Neptune TOI-270~d, \citet{nixon2025} reproduced the JWST abundances of H$_2$O, CH$_4$ and CO$_2$ with magma--atmosphere chemistry at an assumed $f\rm{O}_2 = \Delta$IW-2.8. GJ~9827~d, near the IW buffer in our grid, would sit between these cases, but the planets differ in mass, irradiation and model physics, so this is a comparison rather than a trend. For comparison with Solar System bodies, the average modern terrestrial upper mantle lies about 3.5 log units above IW \citep[e.g.][]{sossi2020}, Solar System bodies span roughly $-2$ to above $+3$ \citep{wadhwa2001, cartier2019}, and polluted white dwarfs indicate that the rocky material they accrete is commonly, but not universally, oxidised \citep{doyle2019a, doyle2019b,Bonsor2023NatAs}. A near-IW mantle combined with of order 1~wt\% of hydrogen ($\ge 6325$~ppmw of the mantle) is broadly consistent with a planet that accreted water-bearing material and equilibrated with a hydrogen-rich primordial envelope while molten, but our grid cannot distinguish this from other pathways that produce the same bulk inventory, and formation and equilibrium models differ on how much water such interactions can produce and retain \citep{werlen2025sub, kimura2026}. If the methane limit holds, this inventory must also be poor in carbon (Sect.~\ref{subsec:disc-methane}), which argues against an origin dominated by carbon-rich solids unless the carbon was sequestered elsewhere \citep{werlen2025co}. Equilibrated atmospheres of planets formed inside the water ice line are predicted to be enriched in H$_2$O and depleted in carbon-bearing species \citep{werlen2026formation}, which would be consistent with this picture. Because the conclusion rests mostly on the two $f\rm{O}_2$ values bracketing the IW buffer, a finer redox grid between $\Delta$IW-1 and $\Delta$IW+2 would sharpen it.

\subsection{Limitations}
\label{subsec:disc-limits}

Several simplifications affect our interpretations above, and some bias them in a known direction.

\emph{Degassing at solidification.} Volatiles are not partitioned into the crystallising solid, and the dissolved inventory is released at once when the melt fraction reaches zero. All 91 runs that reach the present age with a fully solid mantle hold no water in the mantle at 5.465~Gyr. Retention of volatiles in the solid, through mineral solubility or the trapping of interstitial melt during crystallisation \citep{sim2024}, would reduce the late outgassing pulse and the atmospheric inventory of these runs. This affects in particular the 15 fully solid consistent runs, and it may also affect the timing and magnitude of reflation, which in our grid coincides with mantle crystallisation.

\emph{Atmospheric physics.} The atmosphere is ideal-gas and well-mixed, without photochemistry or disequilibrium chemistry. At the $10^3$--$10^5$~bar base pressures of the surviving envelopes, non-ideal behaviour and miscibility between water, hydrogen and silicate could be present \citep{benneke2024jwst, werlen2025sub}, but are not considered. Global equilibrium calculations suggest that non-ideal mixing changes the surface pressure and water content of sub-Neptunes typically by less than 20\% and at most by a factor of two \citep{werlen2026nonideal}, whereas fugacity corrections at GPa pressures can strongly enhance volatile solubility \citep{bower2022, bower2025}. The ideal-gas outgassing thermodynamics used here is considered valid below about $10^5$~bar \citep{bower2025, lichtenberg2026shrink}.

\emph{Escape.} The energy-limited prescription with a fixed efficiency does not fractionate the outflow. Preferential loss of hydrogen relative to oxygen would oxidise the atmosphere over time and raise its water fraction, which would favour slightly more reduced mantles than inferred here. \citet{cesario2026} found the magnitude and timing of reflation to depend on the escape efficiency, so the reflation results here are specific to our adopted efficiency of 0.1. Much lower efficiencies have been argued for heavy, CO$_2$- or SO$_2$-rich atmospheres \citep{postolec2026}. These are less relevant to the H$_2$-bearing envelopes of the consistent runs, but would extend the survival of the oxidised runs. Escape fractionation is a clear future development avenue within the PROTEUS framework \citep{Postolec2026arXiv,Attia2026arXiv}.

\emph{Stellar input and orbit.} The stellar spectrum is a scaled stand-in (HD~85512) rather than a spectrum of GJ~9827, and both tidal heating and orbital evolution are neglected. The XUV history sets the escape timescales of Table~\ref{tab:redox} directly, so the boundary between surviving and stripped runs in hydrogen inventory is uncertain by at least the uncertainty in the integrated XUV fluence.

\emph{Grid coverage.} The three failed cases lie in the reduced, carbon-rich corner and would not have been consistent with the observations. C/H values above unity, which would make carbon a major atmospheric component, are not covered by this grid, and condensed carbon phases are not included in the outgassing calculation.

\section{Conclusions}
\label{sec:conclusions}

We have run a simulation grid of 945 coupled interior--atmosphere evolution models of GJ~9827~d with \textsc{PROTEUS}, varying the hydrogen inventory, the C/H ratio, the core mass fraction and the mantle oxygen fugacity. We compared the end states at 5.465~Gyr with the measured bulk density \citep{passegger2024compact}, the JWST/NIRISS lower limit on atmospheric water and, as a weaker third constraint, the upper limit on methane \citep{piaulet2024jwst}. Our main conclusions are:

\begin{enumerate}

\item \textbf{A magma-ocean origin can reproduce the density and water constraints only in a narrow part of the parameter space.} Of 942 completed runs, 68 are consistent with both. All have a mantle at $f\rm{O}_2 = \Delta$IW$\pm0$ (59 runs) or $\Delta$IW+1 (9 runs) and an initial hydrogen inventory of at least 6325~ppmw of the mantle mass (60-70 Earth oceans).

\item \textbf{Mantle redox controls both atmospheric survival and composition.} Reduced mantles ($f\rm{O}_2 \le \Delta$IW-2) outgas H$_2$--CH$_4$ envelopes, of which 299 of 311 are stripped, most within $10^8$~yr. Oxidised mantles ($f\rm{O}_2 \ge \Delta$IW+2) keep compact H$_2$O-dominated envelopes (89--97\% H$_2$O) that are too dense for GJ~9827~d.

\item \textbf{The atmosphere consistent with observations is mixed, not pure steam.} The consistent runs average 52\% H$_2$O, 45\% H$_2$ and 2\% CH$_4$ by volume, with mean molecular weights of 9.6--14.6~g\,mol$^{-1}$, consistent with the H$_2$/He-background retrieval of JWST/NIRISS ($9.8\pm4.6$~g\,mol$^{-1}$) but below the $\approx$18~g\,mol$^{-1}$ of the agnostic-background retrieval \citep{piaulet2024jwst}.

\item \textbf{Methane constrains the carbon inventory, which density and water do not.} The CH$_4$ VMR of the consistent runs scales nearly linearly with the initial C/H. The JWST upper limit is met by all 38 consistent runs with C/H~$\le 10^{-3}$ and by none with C/H~$\ge 10^{-2}$, which implies C/H~$\lesssim 10^{-3}$--$10^{-2}$ by mass if CH$_4$ is not depleted above the surface. The core mass fraction remains unconstrained.

\item \textbf{Reflation can mimic a low present-day density. The atmospheric methane abundance can decide this evolution path for GJ~9827~d.} Twenty-five of the 110 runs that pass through the measured density and age reflate \citep{cesario2026}, all at $f\rm{O}_2 \le \Delta$IW$\pm0$ and almost all at C/H~$=1$, when a CO-bearing atmosphere is replaced by lighter H$_2$- and CH$_4$-rich gas. Eight consistent runs are reflated at the present age, but their 12--19\% CH$_4$ exceeds the JWST limit.

\item \textbf{Most consistent cases retain a partially molten mantle at 5.5~Gyr} (53 of 68, with melt fractions up to 0.31), held up by the greenhouse effect of a $10^3$--$10^5$~bar envelope, and store up to 61\% of their water in the melt.

\end{enumerate}

The most direct next steps are to compare synthetic transmission spectra of these runs, including photochemistry, directly with the NIRISS and upcoming NIRSpec data, and to include volatile retention in the crystallising mantle and fractionating escape. The approach of filtering a grid of coupled evolution models by observables is directly applicable to other sub-Neptunes with JWST spectra.

\begin{acknowledgements}
I.K. and T.L. acknowledge funding by the European Union (ERC, MagmaWorlds, 101219807). Views and opinions expressed are however those of the author(s) only and do not necessarily reflect those of the European Union or the European Research Council. Neither the European Union nor the granting authority can be held responsible for them. This project was further supported by the Branco Weiss Foundation, the Alfred P. Sloan Foundation (AEThER project, G202114194), NASA’s Nexus for Exoplanet System Science research coordination network (Alien Earths project, 80NSSC21K0593), and the NWO NWA-ORC PRELIFE Consortium (PRELIFE project, NWA.1630.23.013).
H.N. acknowledges support from STFC grant UKRI1184.
C.P.-G. acknowledges support from the E. Margaret Burbidge Prize Postdoctoral Fellowship from the Brinson Foundation, from the Suzuki Postdoctoral Fellowship, and from the Heising-Simons Foundation through the 51 Pegasi b Fellowship program.
The authors thank the Center for Information Technology of the University of Groningen for their support and for providing access to the H\'abr\'ok high-performance computing cluster, on which all simulations presented in this work were performed.
We acknowledge the use of A.I. tools (Claude, GitHub Copilot) for assistance with coding and refinement of the prose. All analyses have been thoroughly vetted by the authors.
\end{acknowledgements}

\bibliographystyle{aa}
\bibliography{ref}

\end{document}